\documentclass[reprint]{JASA}
\usepackage{empheq}
\usepackage{graphicx}
\usepackage{epstopdf, epsfig}
\usepackage{bm}
\usepackage{subfigure}

\begin{document}
%% the square bracket argument will send term to running head in
%% preprint, or running foot in reprint style.

\title[Jet acoustic eigenmodes in the vortex sheet]{Including acoustic modes in the vortex-sheet eigenbasis of a jet}

% ie
%\title[JASA/Sample JASA Article]{Sample JASA Article}

%% repeat as needed
\author{Matteo Mancinelli}\email{matteo.mancinelli@univ-poitiers.fr}\affiliation{D\'{e}partement Fluides Thermique et Combustion, Institut PPRIME - CNRS-Universit\'{e} de Poitiers-ENSMA, 11 Boulevard Marie et Pierre Curie, 86962 Chasseneuil-du-Poitou, Poitiers, France}
\author{Eduardo Martini}\altaffiliation{Also at: Direction des Applications Militaires, CEA-Cesta, 15 Avenue des Sablières, 33114 Le Barp, France}\affiliation{D\'{e}partement Fluides Thermique et Combustion, Institut PPRIME - CNRS-Universit\'{e} de Poitiers-ENSMA, 11 Boulevard Marie et Pierre Curie, 86962 Chasseneuil-du-Poitou, Poitiers, France}
\author{Vincent Jaunet}\affiliation{D\'{e}partement Fluides Thermique et Combustion, Institut PPRIME - CNRS-Universit\'{e} de Poitiers-ENSMA, 11 Boulevard Marie et Pierre Curie, 86962 Chasseneuil-du-Poitou, Poitiers, France}
\author{Peter Jordan}\affiliation{D\'{e}partement Fluides Thermique et Combustion, Institut PPRIME - CNRS-Universit\'{e} de Poitiers-ENSMA, 11 Boulevard Marie et Pierre Curie, 86962 Chasseneuil-du-Poitou, Poitiers, France}

% ie
%\author{Author One}
%\author{Author Two}
%\author{Author Three}

% ie
%\affiliation{Department1,  University1, City, State ZipCode, Country}

%\altaffiliation{}

% may be added after \author{}, ie
% \altaffiliation{Also at: Department1,  University1, City, State ZipCode, Country.}

%% for corresponding author

%% For preprint only,
%  optional, if you want want this message to appear in upper right corner of title page
% \preprint{}

%ie
\preprint{Matteo Mancinelli \textit{et al.}, JASA}

\begin{abstract}
Vortex-sheet models of jets are widely used to describe the dynamics of modes, such as the Kelvin-Helmholtz instability and guided acoustic waves. However, it is seldom pointed out in the literature the absence of the free-stream acoustic modes in the vortex-sheet spectrum. This indicates that free-stream sound waves are not eigensolutions of the parallel jet. This family of modes is important if, for example, one is interested in problems of sound emission or flow-acoustic interactions. In this work we show how a distantly-confined jet may be used as a surrogate problem for the free jet, in which free-stream acoustic waves appear as a set of discrete modes. Comparing the modes observed in the free jet with those of the distantly-confined jet, we show that, other than the free-stream acoustic modes, the eigenvectors and eigenvalues converge with wall distance. The proposed surrogate problem thus efficiently reproduces the dynamics of the original problem, while allowing to account for the dynamics of free-stream acoustic modes.
\end{abstract}

\maketitle

%-----------------------------------------------------------------------------------------------------------------

\section{Introduction}
\label{sec:intro}
The Vortex-Sheet (V-S) model is a simplified inviscid idealisation of a jet where an infinitely thin vortex sheet separates the interior flow from the outer quiescent fluid. The V-S model was first derived by \citet{misc1944landau} and has been largely used since then to: model the jet dynamics and study its stability properties \citep{lessen1965inviscid, michalke1970note}; describe guided jet modes \citep{tam1989three, towne2017acoustic} and the weak, forced resonance they can underpin in free, subsonic jets for jet Mach numbers $0.82\leq M_j<1$ \citep{towne2017acoustic}; model resonance generated in subsonic and supersonic impinging jets \citep{bogey2017feedback, tam1990theoretical}, in a jet-flap interaction configuration \citep{jordan2018jet} as well as in free, supersonic screeching jets \citep{mancinelli2019screech, mancinelli2019reflection, mancinelli2021complex}. The spatial stability properties of a supersonic vortex sheet inside a circular duct in a co-flow configuration were studied as a function of flow and geometric parameters by \citet{chang1993instability}. A double vortex-sheet model was used by \citet{martini_cavalieri_jordan_2019} to further explore the dynamics of guided modes in jets/wakes and the role they play in the onset of absolute instability and their relation to supersonic unstable modes \citep{tam1989three}. \textcolor{red}{The double V-S model was also used to study the effect of confinement on the spatio-temporal stability properties of planar and non-swirling round jets/wakes \citep{juniper2006effect, juniper2008effect}, showing that confinement, under given conditions of density and velocity ratios between the flows and shear-layer thicknesses, may lead to absolute instability of the flow. The stability properties of confined jets reproduce those of free jets for sufficiently large wall distances for waves whose radial support do not vanish slowly in the cross-stream direction \citep{juniper2007full}}.

Its wide-spread use shows that the V-S model is very useful for the analysis and modelling of a variety of shear flows. However, where the free jet is concerned, the V-S eigenspectrum does not include free-stream acoustics. Given that free-stream acoustics  play an important role in jet noise, e.g., in the dynamics of the jet underpinning the flapping modes in supersonic screech \citep{shen2002three, edgington2019aeroacoustic} or resonance in impinging jets \citep{jaunet2019dynamics, edgington2019aeroacoustic}, this absence of free-stream acoustics in the eigenbasis of the free jet is a considerable limitation.

\textcolor{red}{The absence of a subset of eigensolutions in the stability problem was already reported by \citet{gustavsson1979initial} and \citet{ashpis1990vibrating} in the case of boundary-layer flows, showing that waves not forming eigenmodes contribute to the solutions of the associated initial-value problem in the form of a continuous spectrum. This contribution comes out of the deformation of the integration contour in the vicinity of the branch cut in the complex plane. A similar behaviour was also reported by \citet{case1960stability} for an inviscid plane Couette flow for waves not satisfying the eigenvalue problem plus boundary conditions.}

\textcolor{red}{We here show that boundedness condition at infinity, typically imposed in free-jet problems, leads to the absence of free-stream acoustic waves in the eigensolutions of the unconfined V-S model. We then present a strategy that allows free-stream acoustics to be included in an eigen-description of the free jet by avoiding the bounded condition at infinity. Specifically, we propose a surrogate problem involving a distantly-confined jet. By distantly confined we mean that the boundary is located sufficiently far from the vortex sheet such that its influence on the system dynamics is negligible. We show that, besides discrete free-stream acoustic modes, the discrete modes of the free jet are recovered. A model is thereby obtained in which both the usual free-jet eigenmodes are obtained along with a subspace associated with free-stream acoustics.}

The paper is organised as follows. In \S\ref{sec:model} we recall the vortex-sheet model usually adopted in the literature for free jets and present the derivation of the V-S dispersion relation for confined jets which includes free-stream acoustic modes. Results are presented in terms of eigenspectra, eigenfunctions and errors between the free-jet modes and their corresponding modes in the confined jet. Conclusions are finally discussed in \S\ref{sec:conclusions}.

\section{Vortex-sheet model}
\label{sec:model}
The jet is modelled as a cylindrical vortex sheet. All variables are normalised by the cylinder diameter $D$ and ambient density and speed of sound, $\rho_\infty$ and $c_\infty$, respectively. \textcolor{red}{Assuming a parallel base flow $\overline{q}$, the Reynolds decomposition,}

\begin{equation}
\textcolor{red}{q\left(x,r,\theta,t\right)=\overline{q}\left(r\right)+q'\left(x,r,\theta,t\right)\mathrm{,}}
\end{equation}

\noindent \textcolor{red}{is applied to decompose the flow-state vector $q$ into its mean and fluctuating component $\overline{q}$ and $q'$, respectively.} The compressible linearised Euler equations reduce to two convected wave equations for the pressure inside and outside the jet,

\begin{subequations}
\begin{align}
& \left(\left(\frac{\partial}{\partial t}+M_a\frac{\partial}{\partial x}\right)^2 - T\nabla^2\right) p_i'=0\mathrm{,}\label{eq:govern_in}\\
& \left(\frac{\partial^2}{\partial t^2}-\nabla^2\right)p_o'=0\mathrm{,}\label{eq:govern_out}
\end{align}
\label{eq:govern}
\end{subequations}

\noindent where the subscripts $i$ and $o$ denote the interior and outer flows, respectively, $T=T_j/T_\infty$ is the jet-to-ambient temperature ratio and $M_a=U_j/c_\infty$ the acoustic Mach number, with $U_j$ the jet velocity. The relation between the acoustic and jet Mach numbers is $M_j=U_j/c_j=M_a/\sqrt{T}$. We assume the normal-mode ansatz,

\begin{equation}
q'\left(x,r,\theta,t\right)=\hat{q}\left(r\right)e^{i\left(kx+m\theta-\omega t\right)}\mathrm{,}
\end{equation}

\noindent where $k$ is the streamwise wavenumber, $m$ the azimuthal mode and $\omega=2\pi StM_a$ a non-dimensional frequency, with the Strouhal number defined as $St=fD/U_j$ based on the nozzle diameter. Equations \eqref{eq:govern_in} and \eqref{eq:govern_out}, written in cylindrical coordinates, reduce to the modified Bessel equation,

\begin{equation}
\left(\frac{\partial^2}{\partial r^2}+\frac{1}{r}\frac{\partial}{\partial r}-\gamma_{i,o}^2-\frac{m^2}{r^2}\right)\hat{p}_{i,o}=0\mathrm{,}
\label{eq:Bessel}
\end{equation}

\noindent with

\begin{subequations}
\begin{align}
&\gamma_i = \sqrt{k^2-\frac{1}{T}\left(\omega-M_ak\right)^2}\mathrm{,}\label{eq:gamma_i}\\
&\gamma_o = \sqrt{k^2-\omega^2}\mathrm{,}\label{eq:gamma_o}
\end{align}
\end{subequations}

\noindent where the branch cut of the square roots in \eqref{eq:gamma_i} and \eqref{eq:gamma_o} is chosen such that $-\pi/2\leq\mathrm{arg}\left(\gamma_{i,o}\right)<\pi/2$. The solutions of \eqref{eq:Bessel} in each domain are given by,

\begin{subequations}
\begin{empheq}[left=\empheqlbrace]{align}
&\hat{p}_i\left(r\right)=A_iI_m\left(\gamma_ir\right)+B_iK_m\left(\gamma_ir\right)\quad &r\leq 0.5\mathrm{,}\label{eq:sol_general_in}\\
&\hat{p}_o\left(r\right)=A_oI_m\left(\gamma_or\right)+B_oK_m\left(\gamma_or\right)\quad &r>0.5\mathrm{.}\label{eq:sol_general_out}
\end{empheq}
\label{eq:sol_general}
\end{subequations}

These solutions have to be additionally constrained using boundary and matching conditions. Next, we investigate the dispersion relations obtained for free and confined jets and show that the imposition of wave-reflection condition at the wall in the confined case allows to describe free-stream acoustic eigenmodes.

\subsection{Dispersion relation for free jets}
\label{subsec:standard_VS}
We here recall the derivation of the dispersion relation for free jets. To enforce solution amplitudes to be bounded $B_i=0$ and $A_o=0$ are typically set. This is motivated by the fact that for $\gamma_{i,o}\in\mathcal{C}$ $K_m$ and $I_m$ tend to infinity when $r\to 0$ and $r\to\infty$, respectively. The solutions in \eqref{eq:sol_general} become,

\begin{subequations}
\begin{empheq}[left=\empheqlbrace]{align}
&\hat{p}_i\left(r\right)=A_iI_m\left(\gamma_ir\right)\qquad &r\leq 0.5\mathrm{,}\label{eq:sol_in_standard}\\
&\hat{p}_o\left(r\right)=B_oK_m\left(\gamma_or\right)\qquad &r>0.5\mathrm{.}\label{eq:sol_out_standard}
\end{empheq}
\end{subequations}

Note that $\gamma_i$ and $\gamma_o$ play the role of radial wavenumbers in the inner and outer flows, respectively, and the functions $I_m$ and $K_m$ represent incoming and outgoing waves, respectively.

At the vortex sheet, continuity of pressure and vortex-sheet displacement,

\begin{subequations}
\begin{empheq}[left=\empheqlbrace]{align}
& \hat{p}_i\left(\frac{1}{2}\right)=\hat{p}_o\left(\frac{1}{2}\right)\mathrm{,}\label{eq:match_p}\\
& \frac{\partial \hat{p}_i}{\partial r}\left(\frac{1}{2}\right)=\frac{\left(\omega -kM_a\right)^2}{T\omega^2}\frac{\partial \hat{p}_o}{\partial r}\left(\frac{1}{2}\right)\label{eq:match_dp}\mathrm{,}
\end{empheq}
\label{eq:match}
\end{subequations}

\noindent leads to the dispersion relation $D_f\left(k,\omega;M_a,T,m\right)=0$:

{\small
\begin{equation}
\frac{1}{\left(1-\frac{kM_a}{\omega}\right)^2} + \frac{1}{T}\frac{I_m\left(\frac{\gamma_i}{2}\right)\left(\frac{\gamma_o}{2}K_{m-1}\left(\frac{\gamma_o}{2}\right) + mK_m\left(\frac{\gamma_o}{2}\right)\right)}{K_m\left(\frac{\gamma_o}{2}\right)\left(\frac{\gamma_i}{2}I_{m-1}\left(\frac{\gamma_i}{2}\right) - mI_m\left(\frac{\gamma_i}{2}\right)\right)}=0\mathrm{.}
\label{eq:dispersion_standard}
\end{equation}}

Frequency/wavenumber pairs $k$ and $\omega$ define vortex-sheet eigenmodes for given values of azimuthal mode $m$ and flow conditions $M_a$ and $T$. To find these modes, we specify either a real or complex frequency $\omega$ and compute the associated eigenvalue $k$ which solves \eqref{eq:dispersion_standard}. The sign of the group velocity of each mode, which reveals whether a mode travels upstream or downstream, is obtained using the Briggs-Bers criterion \citep{briggs1964electron, bers1983space} and looking at the asymptotic behaviour of $k\left(\omega\right)$ for $\omega_i\to\infty$ \citep{towne2017acoustic}. The wave is downstream-travelling if

\begin{equation}
\lim_{\omega_i\to\infty}k_i=+\infty
\end{equation}

\noindent and upstream-travelling if

\begin{equation}
\lim_{\omega_i\to\infty}k_i=-\infty\mathrm{.}
\end{equation}

\begin{figure}
\centering
\includegraphics[scale=0.17]{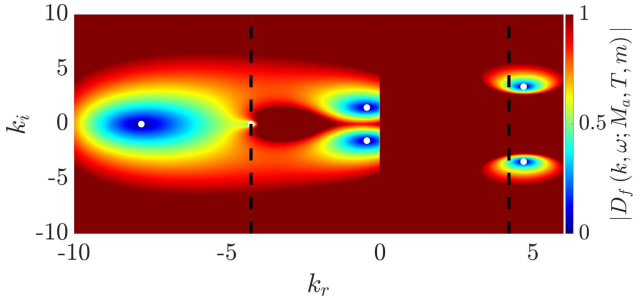}
\caption{Absolute value of the free-jet dispersion relation $D_f\left(k,\omega;M_a,T,m\right)$ in the $k_r$-$k_i$ plane with $\omega\in\mathcal{R}$ for jet Mach number $M_j=1.1$, azimuthal mode $m=0$, Strouhal number $St=0.68$ and jet-to-ambient temperature ratio $T\approx 0.81$. Dashed black lines refer to the ambient speed of sound $\pm c_\infty$, white dots to zeros of the dispersion relation.}
\label{fig:Df}
\end{figure}

Downstream- and upstream-travelling modes are hereafter indicated by a superscript $+$ and $-$, respectively. To calculate the eigenvalues, we first do a coarse grid search for local minima of $D_f\left(k,\omega;M_a,T,m\right)$ in the $k_r$-$k_i$ plane, that is later refined using a root-finder algorithm based on the Levenberg-Marquardt method \citep{levenberg1944method, marquardt1963algorithm} to find zeros of $D_f\left(k,\omega;M_a,T,m\right)$. In what follows we consider the following flow parameters: jet Mach number $M_j=1.1$, azimuthal mode $m=0$, Strouhal number $St=0.68$ and jet-to-ambient temperature ratio $T\approx 0.81$. \textcolor{red}{We show in appendices \ref{app:Mach} and \ref{app:frequency} that the trends and conclusions presented herein are independent of the jet Mach number and frequency considered, respectively.} The absolute value of $D_f\left(k,\omega;M_a,T,m\right)$ in the $k_r$-$k_i$ plane for $\omega\in\mathcal{R}$ is shown in figure \ref{fig:Df} for the flow conditions mentioned above.

\subsubsection{Free-jet eigensolutions}
\label{subsubsec:results_standard}
\begin{figure}
\centering
\includegraphics[scale=0.17]{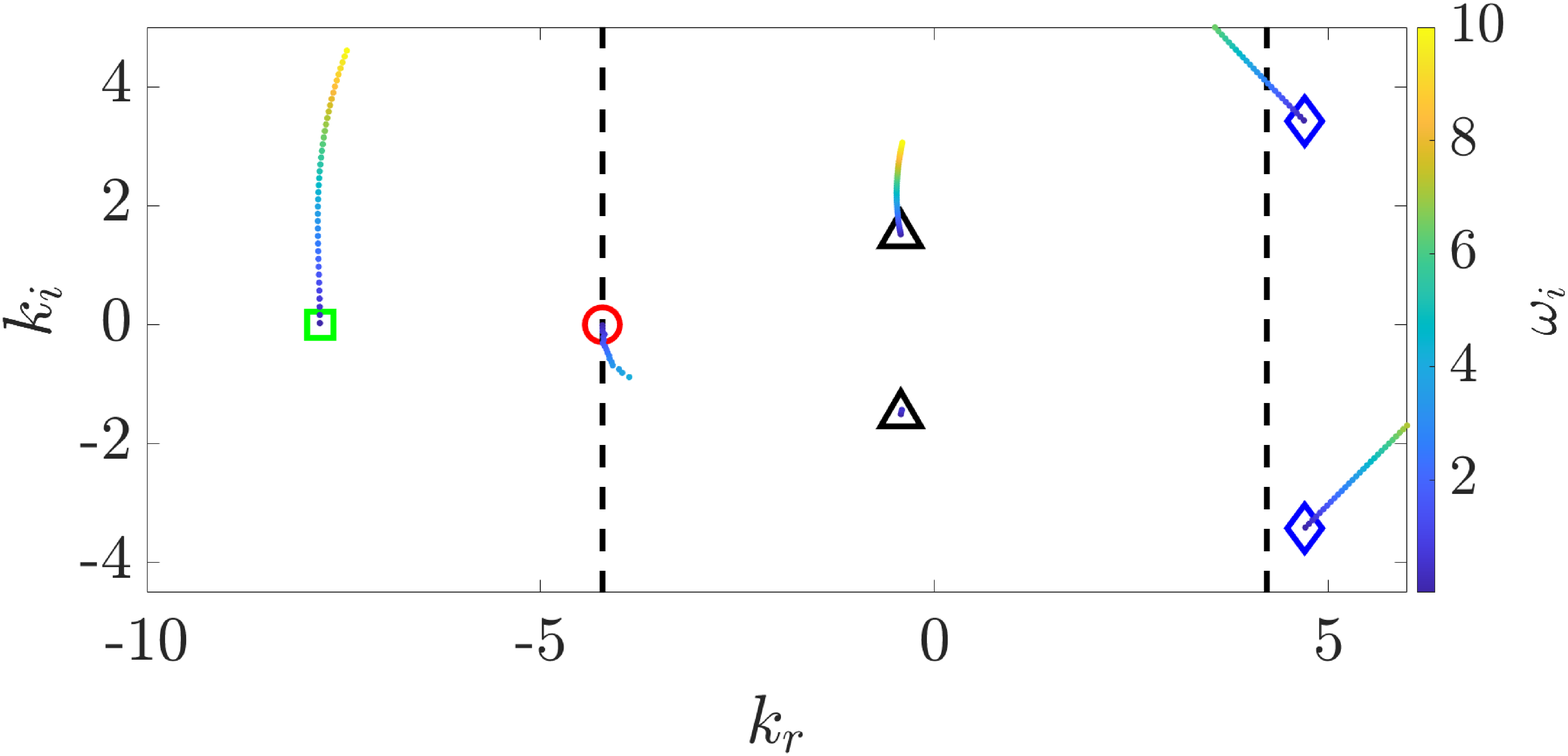}
\caption{Eigenspectrum for $M_j=1.1$, $T\approx 0.81$, $m=0$ and $St=0.68$ obtained using the dispersion relation for free jets \eqref{eq:dispersion_standard}. Empty markers refer to eigenvalues for $\omega\in\mathcal{R}$, $\bullet$ to $\omega\in\mathcal{C}$ with $\omega_i\to\infty$ from blue to yellow, dashed black lines to the ambient speed of sound $\pm c_\infty$. Blue $\diamond$ refer to $k_{KH}^+$ and its complex conjugate, black $\bigtriangleup$ to evanescent $k_p^{\pm}$ for $n=1$, red $\circ$ to propagative $k_p^-$ for $n=2$, green $\Box$ to $k_p^+$ for $n=2$.}
\label{fig:eigenspectrum_standard}
\end{figure}

Figure \ref{fig:eigenspectrum_standard} shows the eigenspectrum in the $k_r$-$k_i$ plane for the flow parameters listed above. Lines with phase speed equal to $\pm c_\infty$ are shown to identify modes that are supersonic with respect to the free stream. Eigenvalues are calculated for both real and complex values of $\omega$. Distinct families of modes can be identified. The V-S model supports one convectively unstable mode, that is the Kelvin-Helmholtz (K-H) mode, which is hereafter denoted $k_{KH}^+$, and its complex conjugate. The vortex sheet also supports guided modes, which are hereafter denoted $k_p$. These modes belong to a hierarchical family of modes identified by the azimuthal and radial orders $m$ and $n$, respectively, and are propagative only in a well-defined $St$-number range which is delimited by a branch point and a saddle point \citep{tam1989three}. Following \citet{towne2017acoustic}, these modes can be completely trapped inside the jet, experiencing it as a soft-walled duct, depending on the $St$ and radial order $n$ considered. \citet{martini_cavalieri_jordan_2019} showed that this behaviour can be explained by the effective impedance of the VS, which for these conditions is close to zero, thus emulating a pressure-release condition. For the flow conditions we consider, we note: the propagative $k_p^-$ with $n=2$, which is characterised by a subsonic phase speed and lies immediately next to the sonic line; the evanescent $k_p^{\pm}$ for $n=1$, which have supersonic phase speed (mode with $\vert k_i\vert>0$ is downstream-travelling, whereas the mode with $\vert k_i\vert <0$ is upstream-travelling); and the propagative $k_p^+$ with $n=2$, which lies on the real axis and has a subsonic phase speed. As outlined in \S\ref{sec:intro}, free-stream acoustic modes are not found in the eigenspectrum. \textcolor{red}{We remind the reader that this does not imply that free-stream acoustic waves are not solutions of the unconfined vortex sheet. They are. And they come along as solutions in the impulse-response or initial-value problem in the form of a continuous spectrum. It is their presence in the eigenbasis description which appears to be missing.}

\begin{figure}
\centering
\includegraphics[scale=0.18]{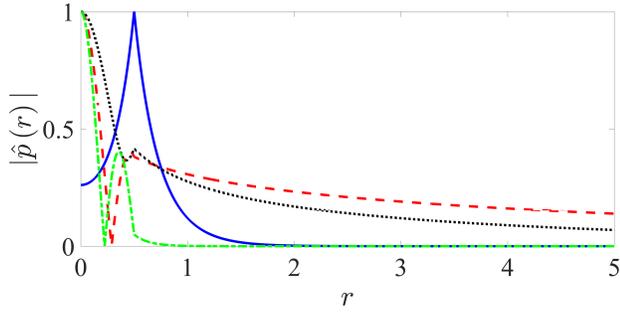}
\caption{Normalised pressure eigenfunctions obtained using the unconfined vortex sheet for jet Mach number $M_j=1.1$, jet-to-ambient temperature ratio $T\approx 0.81$, azimuthal mode $m=0$ and $St=0.68$. Solid blue line refers to $k_{KH}^+$, dashed red line to $k_p^-$ for $n=2$, black dotted line to $k_p^-$ for $n=1$, dash-dotted green line to $k_p^+$ for $n=2$.}
\label{fig:eigenfunctions_standard}
\end{figure}

Figure \ref{fig:eigenfunctions_standard} shows an example of the pressure eigenfunctions, which are normalised in order to have unitary maximum amplitude. For the sake of brevity, we here report only the eigenfunctions of the K-H mode, the upstream-travelling guided modes of first and second radial orders and the downstream-travelling guided mode of second radial order. As expected, the K-H wave shows an eigenfunction with maximum amplitude at the vortex sheet and a fast radial decay. Concerning the guided jet modes, as outlined above, the $k_p^-$ modes for $n=1$ and $2$ show an eigenfunction with a radial support both inside and outside the jet with a slow radial decay, whereas the $k_p^+$ mode for $n=2$ is almost completely trapped in the inner region for the flow conditions considered.

\subsection{Dispersion relation for confined vortex sheet}
\label{subsec:modified_VS}
Considering a confined jet with boundary located at $r=r_{MAX}$, enforcing to the general solution in \eqref{eq:sol_general} bounded conditions at $r=0$ and soft-wall boundary conditions at $r=r_{MAX}$, i.e., $p\left(r_{MAX}\right)=0$, we obtain,

\begin{subequations}
\begin{align}
& B_i=0\mathrm{,}\\
& A_o=-B_o\frac{K_m\left(\gamma_o r_{MAX}\right)}{I_m\left(\gamma_o r_{MAX}\right)}=-B_oz\left(r_{MAX}\right)\mathrm{,}
\end{align}
\end{subequations}

\noindent where for notational simplicity we define $z\left(r_{MAX}\right)=K_m\left(\gamma_o r_{MAX}\right)/I_m\left(\gamma_o r_{MAX}\right)$. The Dirichlet boundary condition used at $r=r_{MAX}$ implies a pressure release condition, i.e., a soft-walled duct, and an in-phase wave reflection. The solution for the pressure in the inner and outer flow becomes,

\begin{subequations}
\begin{empheq}[left=\empheqlbrace]{align}
&\hat{p}_i\left(r\right)=A_iI_m\left(\gamma_ir\right)&r\leq 0.5\\
&\hat{p}_o\left(r\right)=B_o\left(-zI_m\left(\gamma_or\right)+K_m\left(\gamma_or\right)\right)&r>0.5\mathrm{.}
\end{empheq}
\label{eq:sol_generalised}
\end{subequations}

\begin{figure}
\centering
\includegraphics[scale=0.18]{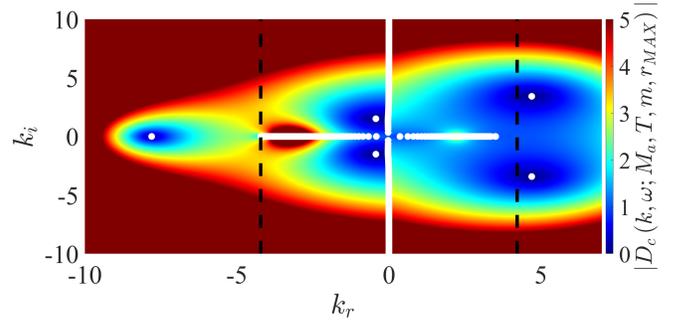}
\caption{Absolute value of the distantly-confined-jet dispersion relation $D_c\left(k,\omega;M_a,T,m,r_{MAX}\right)$ in the $k_r$-$k_i$ plane with $\omega\in\mathcal{R}$ for jet Mach number $M_j=1.1$, azimuthal mode $m=0$, Strouhal number $St=0.68$ and jet-to-ambient temperature ratio $T\approx 0.81$. Dashed black lines refer to the ambient speed of sound $\pm c_\infty$, white dots to zeros of the dispersion relation.}
\label{fig:Dc}
\end{figure}

Imposing the solution match at the vortex-sheet location \eqref{eq:match}, we obtain the dispersion relation $D_c\left(k,\omega;M_a,T,m,r_{MAX}\right)=0$,

{\footnotesize
\begin{equation}
\begin{split}
&\frac{1}{\left(1-\frac{kM_a}{\omega}\right)^2} + \frac{1}{T}\frac{I_m\left(\frac{\gamma_i}{2}\right)}{K_m\left(\frac{\gamma_o}{2}\right)-zI_m\left(\frac{\gamma_o}{2}\right)}\\
&\frac{\frac{\gamma_o}{2}K_{m-1}\left(\frac{\gamma_o}{2}\right) + mK_m\left(\frac{\gamma_o}{2}\right)+z\left(\frac{\gamma_o}{2}I_{m-1}\left(\frac{\gamma_o}{2}\right)-mI_m\left(\frac{\gamma_o}{2}\right)\right)}{\frac{\gamma_i}{2}I_{m-1}\left(\frac{\gamma_i}{2}\right) - mI_m\left(\frac{\gamma_i}{2}\right)}=0\mathrm{,}
\end{split}
\label{eq:dispersion_generalised}
\end{equation}}

\noindent which differs from the dispersion relation used in the unconfined vortex sheet by the additional terms in $z\left(r_{MAX}\right)$. The absolute value of $D_c\left(k,\omega;M_a,T,m,r_{MAX}\right)$ for $r_{MAX}=100$ in the $k_r$-$k_i$ plane for a real frequency $\omega$ is shown in figure \ref{fig:Dc}. In addition to the minima already observed in figure \ref{fig:Df} for the free-jet dispersion relation, a series of local minima on the real and imaginary axes are found within the free-stream supersonic region.

\begin{figure}
\centering
\subfigure[]{\includegraphics[scale=0.17]{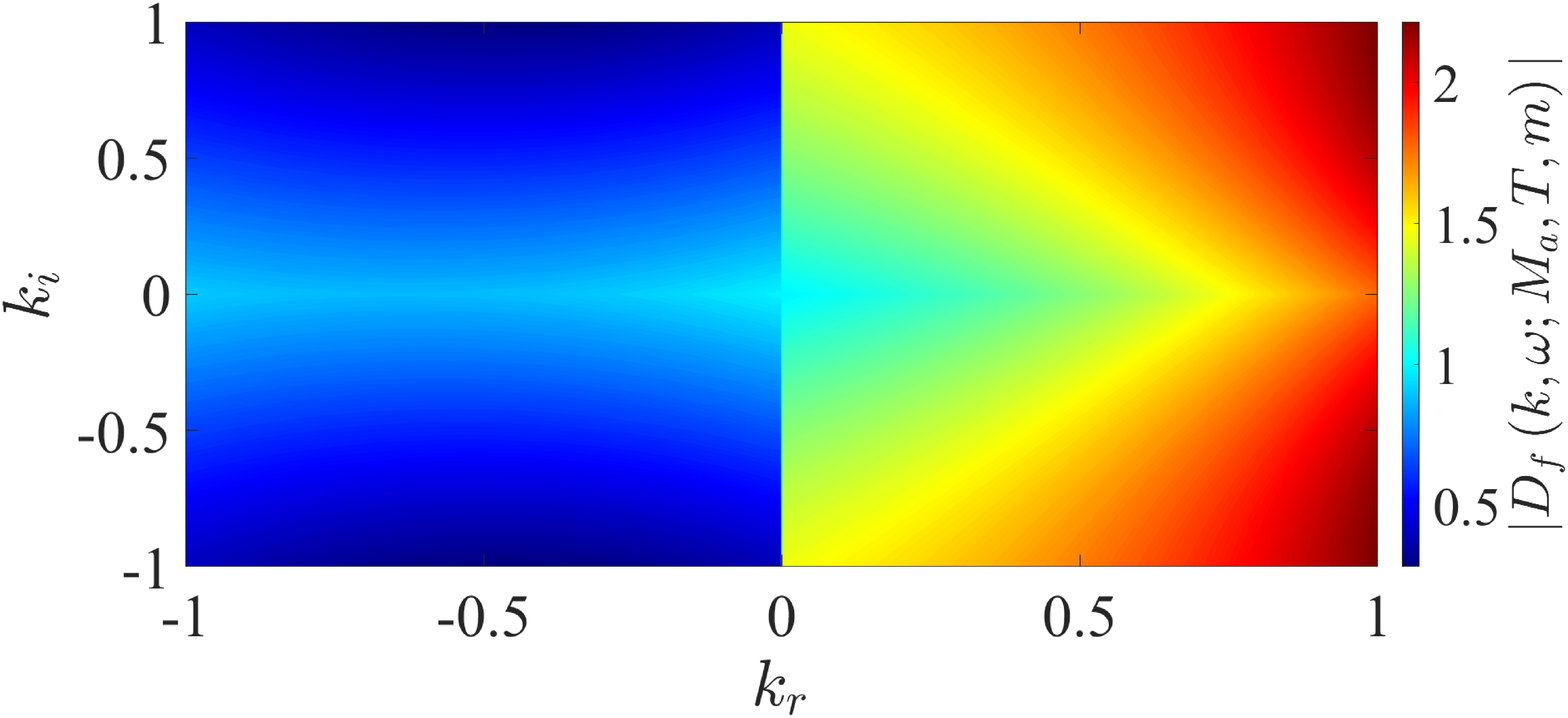}\label{fig:Df_contour}}
\subfigure[]{\includegraphics[scale=0.17]{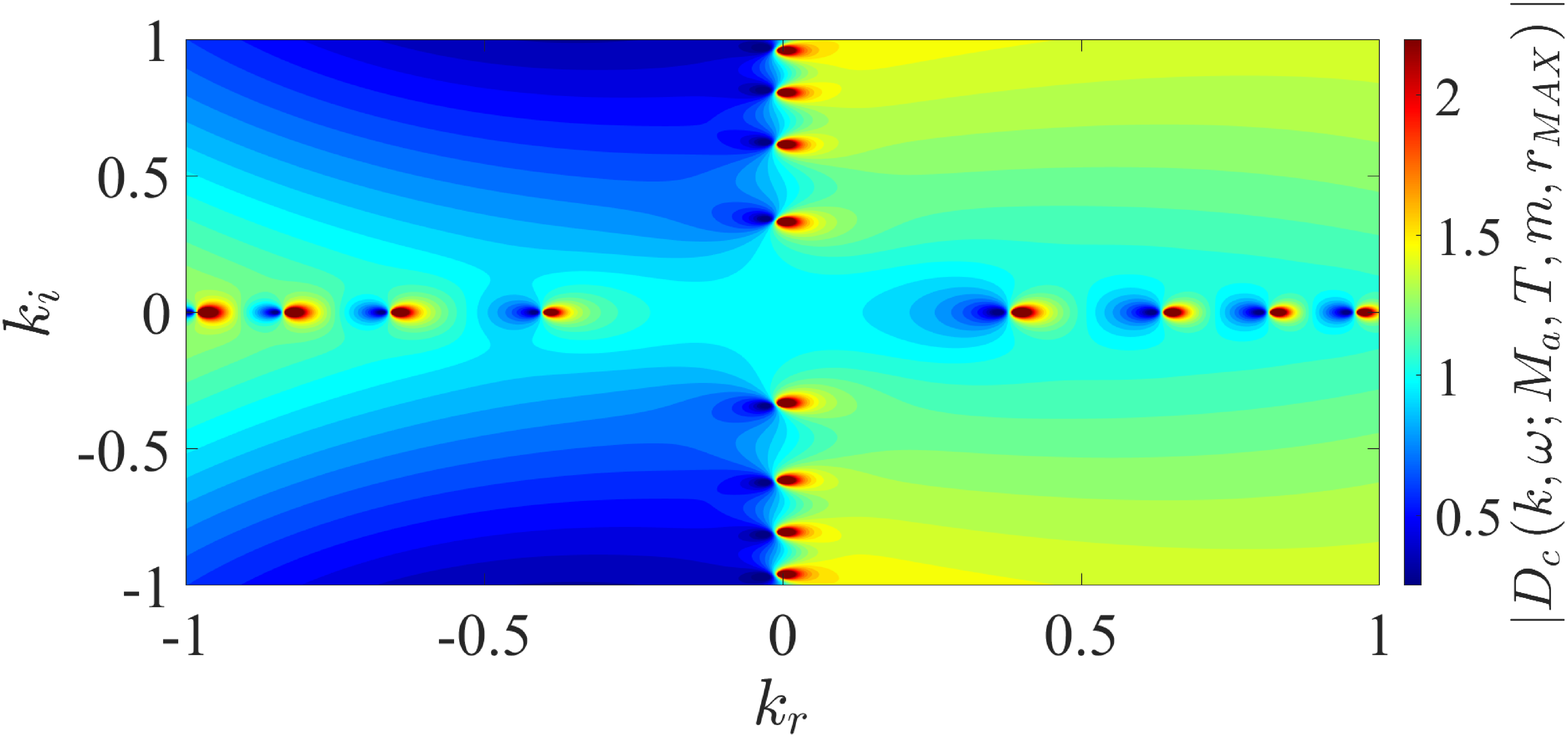}\label{fig:Dc_contour}}
\caption{\textcolor{red}{Absolute value of the dispersion relations in the $k_r$-$k_i$ plane around the origin for $\omega\in\mathcal{R}$, jet Mach number $M_j=1.1$, azimuthal mode $m=0$, Strouhal number $St=0.68$ and jet-to-ambient temperature ratio $T\approx 0.81$. (a) unconfined vortex sheet, (b) confined vortex sheet.}}
\label{fig:D_contour}
\end{figure}

\textcolor{red}{The different behaviour of the dispersion relation between unconfined and confined cases is better highlighted in figure \ref{fig:D_contour} which shows a zoomed portrait of the absolute value of $D_f$ and $D_c$ in the $k_r$-$k_i$ plane around the origin. A branch cut inducing a discontinuity and a discontinuous derivative of the dispersion relation on the imaginary and real axes, respectively, is found in the unconfined case. The contribution of the free acoustics to an impulse response or initial-value problem would appear in the form of a continuous spectrum due to the deformation of the contour integral around the branch cut \citep{case1960stability,gustavsson1979initial,ashpis1990vibrating}. On the contrary, for the confined case, the imposition of the wave-reflection condition at infinity appears to remove the discontinuities along the real and imaginary axes where zeros of the dispersion relation associated with free-stream acoustic eigenmodes are found.}

\subsubsection{Confined-jet eigensolutions}
\label{subsubsec:results_generalised}
\begin{figure}
\centering
\includegraphics[scale=0.17]{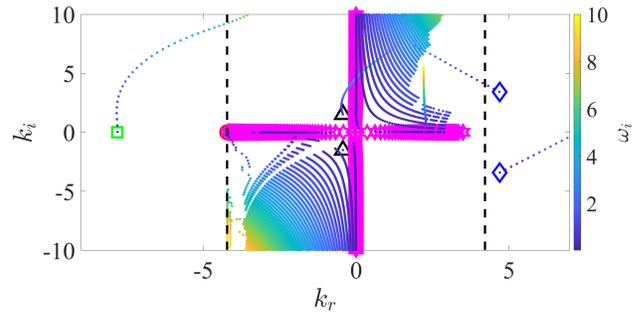}
\caption{Eigenspectrum for $M_j=1.1$, $T\approx 0.81$, $m=0$ and $St=0.68$ obtained using the dispersion relation for confined jets \eqref{eq:dispersion_generalised}. Empty markers refer to eigenvalues for $\omega\in\mathcal{R}$, $\bullet$ to $\omega\in\mathcal{C}$ with $\omega_i\to\infty$ from blue to yellow, dashed black lines to the ambient speed of sound $\pm c_\infty$. Markers and colours are the same used in figure \ref{fig:eigenspectrum_standard} to distinguish the modes and magenta $\star$ refer to $k_a^\pm$.}
\label{fig:eigenspectrum_generalised}
\end{figure}

\begin{figure}
\centering
\includegraphics[scale=0.18]{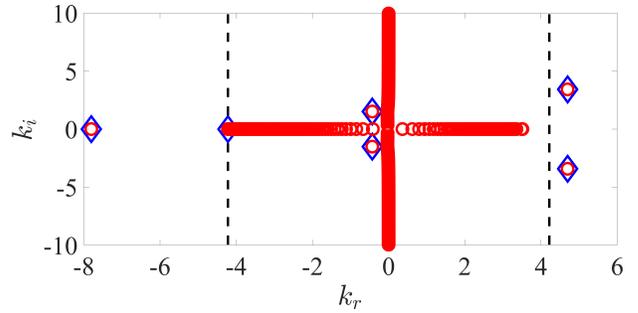}
\caption{Comparison between eigenspectra obtained using the dispersion relation \eqref{eq:dispersion_standard} for free jets (blue $\diamond$) and the dispersion relation for confined jets \eqref{eq:dispersion_generalised} (red $\circ$) using $r_{MAX}=100$.}
\label{fig:eig_comparison}
\end{figure}

Figure \ref{fig:eigenspectrum_generalised} shows the eigenspectrum obtained using the dispersion relation for the confined vortex sheet \eqref{eq:dispersion_generalised}. We here consider a maximum radial distance $r_{MAX}=100$, the flow may thus be considered as distantly confined. Eigenvalues for both real and complex frequencies are represented. We note that, in addition to the modes observed in figure \ref{fig:eigenspectrum_standard} using the unconfined vortex sheet, propagative and evanescent free-stream acoustic modes, that hereafter are denoted $k_a^{\pm}$, are also present. \textcolor{red}{These modes are routinely found in numerical eigensolutions of the finite-thickness jet. But their dependence on the distant boundary condition shown here illustrates how their appearance in the finite-thickness eigensolution as well could likely arise due to the numerical imposition of a distant boundary that is usually implemented to solve the eigenvalue problem.} We note that all the modes found using the dispersion relation \eqref{eq:dispersion_standard} in unconfined conditions are reproduced by the dispersion relation for the distantly confined vortex sheet \eqref{eq:dispersion_generalised}, as shown in figure \ref{fig:eig_comparison} (for the sake of brevity we hereafter show only results for $\omega\in\mathcal{R}$). The distantly-confined problem, thus, accurately describes the dynamics of the free vortex sheet while also exhibiting an additional subspace associated with free-stream acoustics.

\begin{figure}
\centering
\includegraphics[scale=0.18]{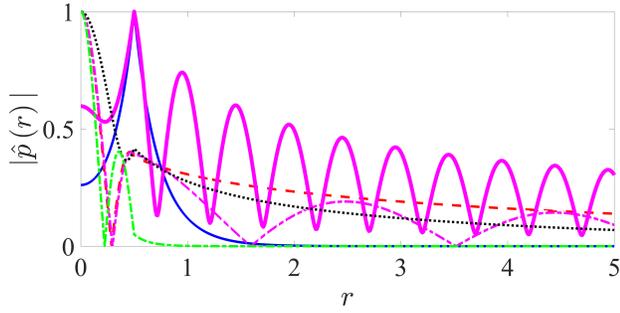}
\caption{Eigenfunctions obtained using the solution in \eqref{eq:sol_generalised} and the dispersion relation for confined jets in \eqref{eq:dispersion_generalised} to find the eigenvalues. Colour lines are the same used in figure \ref{fig:eigenfunctions_standard} for free jets. Additionally, dash-dotted magenta line refers to propagative $k_a^-$ mode, bold magenta line to evanescent $k_a^-$ mode.}
\label{fig:eigenfunctions_generalised}
\end{figure}

Figure \ref{fig:eigenfunctions_generalised} shows the eigenfunctions obtained using the pressure solution in \eqref{eq:sol_generalised} and the dispersion relation in \eqref{eq:dispersion_generalised} for the eigenvalue computation. For the sake of brevity, we only show eigenfunctions for the same modes considered in the unconfined case in figure \ref{fig:eigenfunctions_standard} and additionally one of the propagative and evanescent $k_a^-$ modes. We note that the eigenfunctions for the K-H and the guided jet modes are identical to those of figure \ref{fig:eigenfunctions_standard} obtained using the unconfined vortex-sheet model. The eigenfunctions of the propagative and evanescent $k_a^-$ modes show the expected shape for acoustic modes with a spatial support both inside and outside the jet and an algebraic amplitude decay for increasing $r$.

\subsubsection{Asymptotic behaviour for large wall distances}
\label{subsec:convergence}
\begin{figure}
\centering
\subfigure[]{\includegraphics[scale=0.18]{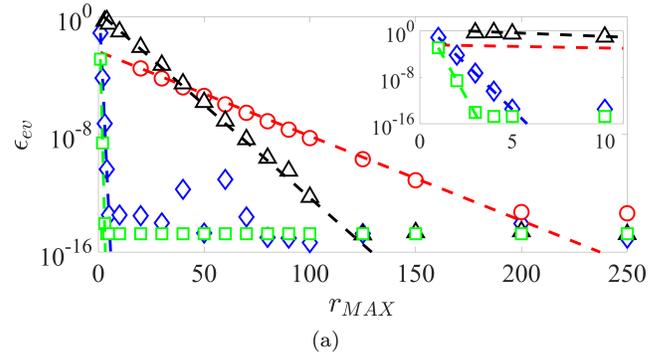}}
\subfigure[]{\includegraphics[scale=0.18]{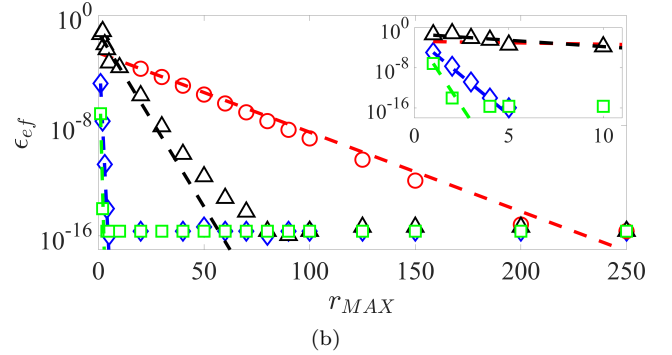}}
\caption{Trend of the eigensolution error between the confined- and free-jet cases as a function of $r_{MAX}$: (a) eigenvalues, (b) eigenfunctions. Markers correspond to errors between eigensolutions obtained from the dispersion relations \eqref{eq:dispersion_generalised} and \eqref{eq:dispersion_standard}, dashed lines to the exponential fit dictated by \eqref{eq:z_infty}. Blue $\diamond$ correspond to the $k_{KH}^+$ mode, red $\circ$ to the $k_p^-$ mode for $n=2$, black $\bigtriangleup$ to the $k_p^-$ mode for $n=1$ and green $\Box$ to the $k_p^+$ for $n=2$. A zoom for $r_{MAX}\leq 11$ is reported at the top right to better show the trend for the $k_{KH}^+$ and $k_p^+$ modes.}
\label{fig:errors}
\end{figure}

We now study the effect of the domain size on the eigensolution computation in the limit of large wall distances. \textcolor{red}{As outlined by \citet{juniper2008effect}, the stability properties of a confined jet are equivalent to those of a free jet for waves whose spatial support decays rapidly in the radial direction if the boundary is located sufficiently far from the flow. In this work we explore values of $r_{MAX}\to\infty$ and thus the stability of the confined and unconfined systems are equivalent. Non-acoustic modes of the confined problem converge to similar solutions for the unconfined problem.} We here quantify this convergence by looking at the discrepancy between eigensolutions computed in the unconfined case and eigensolutions computed in the distantly-confined case for different values of $r_{MAX}$. Specifically, we select the K-H mode, $k_p^\pm$ modes for $n = 2$ and $k_p^-$ mode for $n=1$ and we compute the error between the eigenvalues and the eigenfunctions calculated in the free and distantly-confined cases for increasing values of $r_{MAX}$. The error value for the eigenvalues and the eigenfunctions is defined as follows,

\begin{subequations}
\begin{align}
& \epsilon_{ev}=\vert k_f- k_c\vert\mathrm{,}\label{eq:error_value}\\
& \epsilon_{ef}=1-\bigg|\frac{\langle\hat{p}_f,\hat{p}_c\rangle}{\sqrt{\langle \hat{p}_f,\hat{p}_f\rangle\langle\hat{p}_c,\hat{p}_c\rangle}}\bigg|\mathrm{,}\label{error_function}
\end{align}
\label{eq:error}
\end{subequations}

\noindent where the subscripts f and c indicate solutions from the free- and confined-jet cases, respectively, and the inner product $\langle\cdot,\cdot\rangle$ represents an euclidean dot product. Figure \ref{fig:errors} shows the trend of the eigensolution error between the confined and unconfined vortex sheet as a function of $r_{MAX}$, which we let vary in the range $\left[1,250\right]$. For both eigenvalues and eigenfunctions the confined-jet solution matches the free-jet solution (up to machine precision) as $r_{MAX}$ becomes sufficiently large. We observe that the error trend is different depending on the mode considered. Specifically, the convergence trend is very fast for the K-H wave and the downstream-travelling guided mode of second radial order for which the error value reaches machine precision for $r_{MAX}=5$. \textcolor{red}{We point out that the small error increase for the K-H eigenvalue observed for $40\leq r_{MAX}\leq 70$ is most likely related to numerical inaccuracies in the root-finder algorithm when searching for zeros of the dispersion relation and, anyway, never exceeds the value of $10^{-12}$.} On the contrary, the convergence speed is slower for the $k_p^-$ mode with $n=1$ and, above all, for the $k_p^-$ mode with $n=2$. For this mode the error value becomes lower than $10^{-9}$ for $r_{MAX}\geq 100$, distance at which the confined-jet solution may be considered equal to the free-jet one. We point out that the different convergence rate with $r_{MAX}$ observed for the modes is consistent with their radial support shown in figure \ref{fig:eigenfunctions_generalised} and the conclusions reported by \citet{juniper2007full}. Specifically, for modes with eigenfunctions concentrated either in the core or at the vortex sheet of the jet, such as the $k_p^+$ mode with $n=2$ and the $k_{KH}^+$ wave, the confined-jet solution equals the free-jet one for low values of $r_{MAX}$. On the contrary, modes with larger eigenfunction support outside the jet, such as the $k_p^-$ modes with $n=1$ and $2$, require a larger $r_{MAX}$ value in order for the confined-jet solution to match the free-jet one.

To better understand the convergence trend reported above, we here analyse the behaviour of the above-defined parameter $z\left(r_{MAX}\right)=K_m\left(\gamma_o r_{MAX}\right)/I_m\left(\gamma_o r_{MAX}\right)$ when $r_{MAX}\to\infty$. For this aim, we consider the zero-order asymptotic expansion at infinity of the Bessel functions $I$ and $K$, which, following \citet{abramowitz1948handbook}, can be expressed as,

\begin{subequations}
\begin{equation}
\lim_{r_{MAX}\to\infty}I_m\left(\gamma_o r_{MAX}\right)\approx \frac{e^{\gamma_o r_{MAX}}}{\sqrt{2\pi\gamma_o r_{MAX}}}\mathrm{,}
\label{eq:I_asymptotic}
\end{equation}
\begin{equation}
\lim_{r_{MAX}\to\infty}K_m\left(\gamma_o r_{MAX}\right)\approx\sqrt{\frac{\pi}{2\gamma_o r_{MAX}}}e^{-\gamma_o r_{MAX}}\mathrm{,}
\label{eq:K_asymptotic}
\end{equation}
\end{subequations}

\noindent such that $z$ for $r_{MAX}\to\infty$ can be evaluated as follows,

{\small
\begin{equation}
z\left(r_{MAX}\to\infty\right)=\lim_{r_{MAX}\to\infty}\frac{K_m\left(\gamma_o r_{MAX}\right)}{I_m\left(\gamma_o r_{MAX}\right)}\approx\pi e^{-2\gamma_o r_{MAX}}\mathrm{.}
\label{eq:z_infty}
\end{equation}}

Assuming that the exponent is not purely imaginary, equation \eqref{eq:z_infty} shows that the convergence of the confined-jet solution to the free-jet one should have an exponential trend with the wall distance $r_{MAX}$. This result is confirmed by the comparison of the errors between the eigensolutions obtained in the confined and unconfined cases with the exponential fit dictated by \eqref{eq:z_infty} in figure \ref{fig:errors}. The convergence rate of all the modes is well approximated by the exponential fit until the error reaches a plateau at machine precision.

\begin{figure}
\centering
\subfigure[]{\includegraphics[scale=0.17]{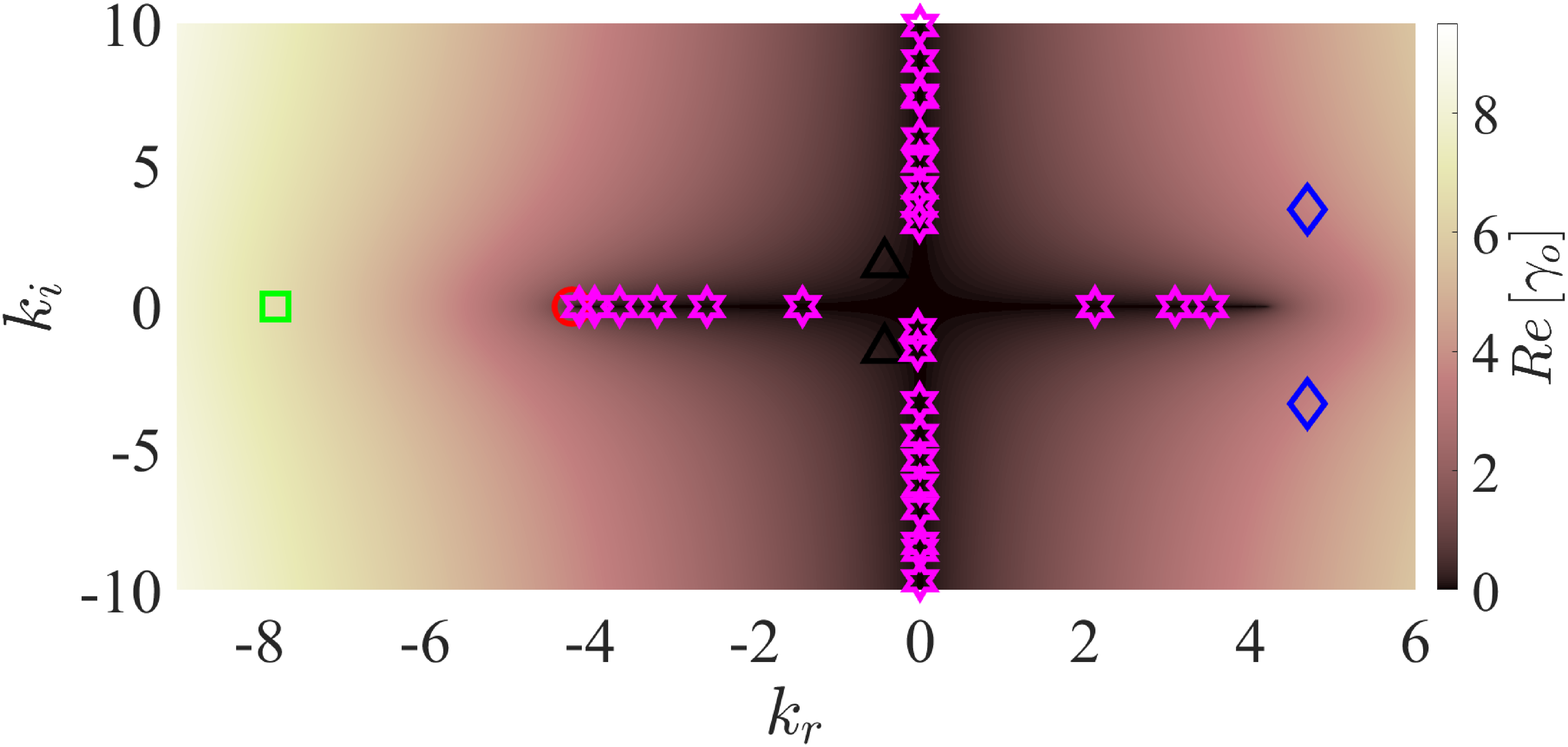}\label{fig:real_gamma_o}}
\subfigure[]{\includegraphics[scale=0.17]{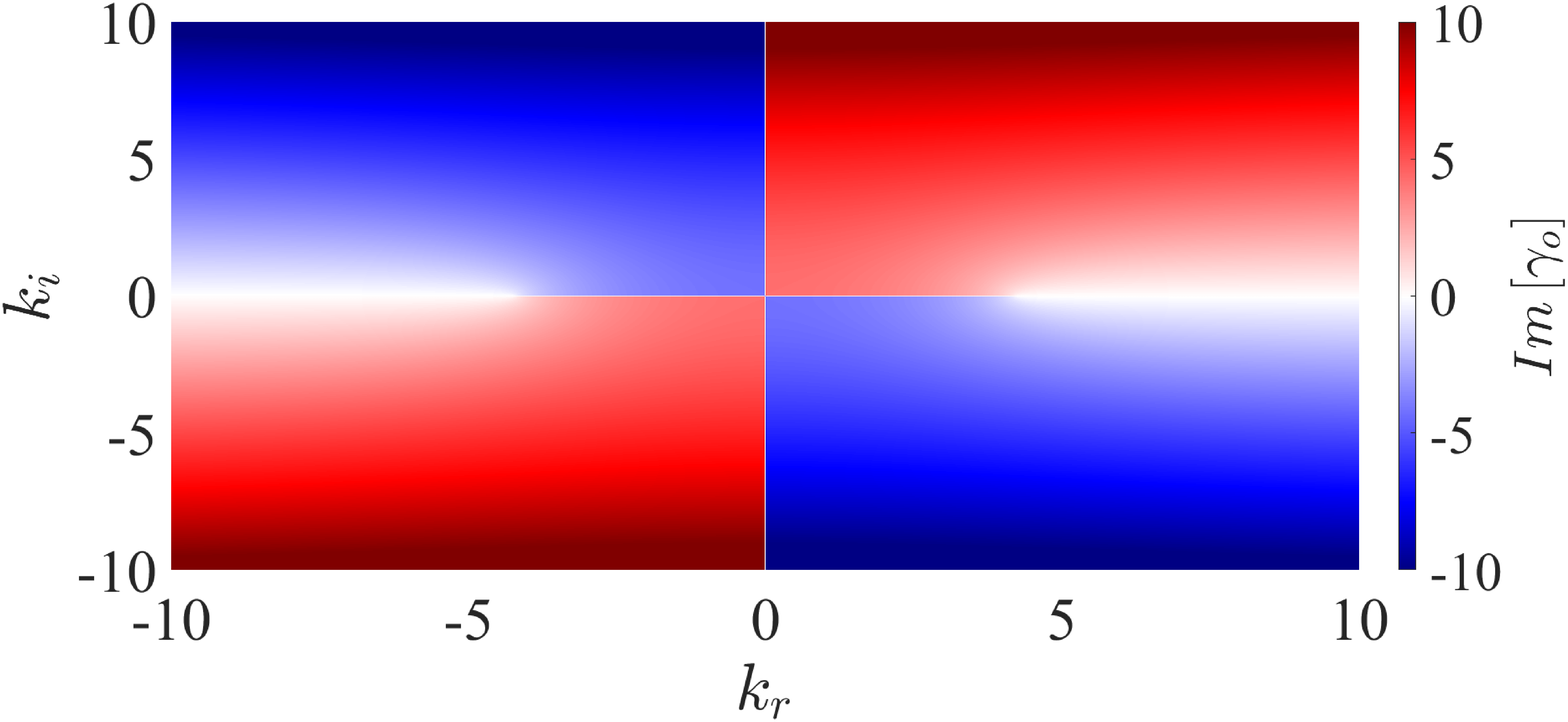}\label{fig:imag_gamma_o}}
\caption{Contour map of $\gamma_o$ \eqref{eq:gamma_o} in the $k_r$-$k_i$ plane: (a) real part of $\gamma_o$ with eigenspectrum for confined jet superimposed on it (for the sake of clarity of representation free-stream acoustic modes are plotted skipping twenty modes between each marker); (b) imaginary part of $\gamma_o$.}
\label{fig:gamma_o}
\end{figure}

\begin{figure}
\centering
\includegraphics[scale=0.18]{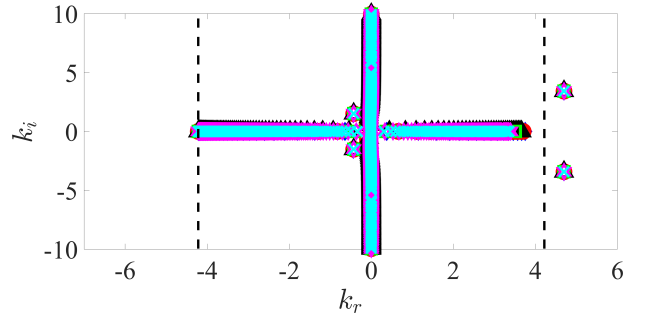}
\caption{Zoom of the eigenspectra of a confined jet for $\omega\in\mathcal{R}$ in the $k_r$-$k_i$ plane for different values of $r_{MAX}$: blue $\diamond$ refer to $r_{MAX}=20$, red $\circ$ to $r_{MAX}=30$, black $\bigtriangleup$ to $r_{MAX}=50$, green $\Box$ to $r_{MAX}=100$, magenta $\ast$ to $r_{MAX}=150$, cyan $\times$ to $r_{MAX}=200$.}
\label{fig:eigenspec_rComparison}
\end{figure}

As mentioned above, the limit in \eqref{eq:z_infty}, and hence the convergence rate of the errors $\epsilon_{ev}$ and $\epsilon_{ef}$, depend on whether $\gamma_o$ \eqref{eq:gamma_o} is complex or purely imaginary. Specifically, when $\gamma_o\in\mathcal{C}$ the limit in \eqref{eq:z_infty} is equal to zero, the dispersion relation $D_c$ asymptotically approaches the free-jet one $D_f$ and the confined vortex-sheet solution converges to the unconfined one for large $r_{MAX}$. The convergence rate depends on how large is the real part of $\gamma_o$, the larger the value of $Re\left[\gamma_o\right]$ the faster the convergence. When $\gamma_o\in\mathcal{I}$, that is when $\vert k\vert < \omega$ or $k\in\mathcal{I}$, the limit in \eqref{eq:z_infty} is undefined and assumes a finite oscillatory value equal to $\pi e^{2i\left[0,\pi\right]}$. Figure \ref{fig:real_gamma_o} shows the contour map of the real part of $\gamma_o$ in the $k_r$-$k_i$ plane. The eigenspectrum of the confined jet for $r_{MAX}=100$ is also superimposed on the plot (note that for the sake of clarity of the representation free-stream acoustic modes are reported on the graph skipping twenty modes between each marker). Consistent with what said above, the real part of $\gamma_o$ associated with the K-H wave and $k_p^+$ mode with $n=2$ is much larger than that related to the $k_p^-$ modes of first and second radial orders, thus explaining the different convergence rate observed in figure \ref{fig:errors}. We also note that the free-stream acoustic modes are characterised by a purely imaginary $\gamma_o$ value and they lie on the branch cut of $Im\left[\gamma_o\right]$ where a phase jump of $\pi$ occurs, as shown in figure \ref{fig:imag_gamma_o}. Given that $\gamma_o\in\mathcal{I}$ for these modes, according to \eqref{eq:z_infty} the eigensolutions do not converge as $r_{MAX}$ increases. This behaviour is exemplified by figure \ref{fig:eigenspec_rComparison}, which shows the eigenspectra in the case of a confined jet for different values of the distance of the wall from the jet, that is $r_{MAX}=20$, $30$, $50$, $100$, $150$ and $200$. To better focus on the effect of the wall distance on the acoustic modes, we here plot a zoom on the free-stream supersonic region. \textcolor{red}{A different discretisation of the continuous branch of the free-stream acoustic modes is obtained for each wall distance. We point out that the grid search in the complex plane used to find zeros of the dispersion relation has to be increasingly refined to capture propagative acoustic waves close to the sonic line for increasing $r_{MAX}$ values. Despite this refinement, near-sonic, downstream-travelling, propagative acoustic eigenmodes are found only for few $r_{MAX}$ values and complex frequencies $\omega$. This behaviour is ascribed to numerical issues when searching for zeros of the dispersion relation.}

\section{Conclusions}
\label{sec:conclusions}
In this paper we dealt with the issue of describing the free-stream acoustic modes of a jet using a vortex-sheet model. Despite the widespread use of the vortex-sheet model for describing jet dynamics, the absence of eigenmodes associated with free-stream acoustics due to the imposition of bounded solution at infinity, seems to have been overlooked. To describe free-stream acoustics we propose a distantly-confined jet as a surrogate problem for a free jet. The surrogate problem asymptotically exhibits all of the modes supported by the free jet in addition to a discrete family of free-stream acoustic modes. When the wall is taken sufficiently far from the vortex sheet, the eigensolutions of the confined jet match those of the free jet, such that free-jet stability properties are fully recovered. Specifically, the discrepancy between the eigensolutions of the confined- and free-jet cases decays exponentially with the wall distance. We show that the convergence rate of the confined-jet eigensolutions depends on the real part of $\gamma_o=\sqrt{k^2-\omega^2}$, which plays the role of a radial wavenumber. Larger values imply a faster radial decay of the eigenfunction, which is consequently less influenced by the wall, and thus a faster convergence to the free-jet eigensolution.

\begin{acknowledgments}
M.M. acknowledges the financial support of CNES (Centre National d'Etudes Spatiales) under a post-doctoral grant at the time of the writing of the manuscript. E.M. acknowledges the support of CEA-CESTA (Commissariat à l'\'{E}nergie Atomique - Centre d'Etudes Scientifiques et Techniques d'Aquitaine) under a post-doctoral grant.
\end{acknowledgments}

\appendix

\section{Jet Mach number effect}
\label{app:Mach}
\begin{figure}
\centering
\subfigure[]{\includegraphics[scale=0.17]{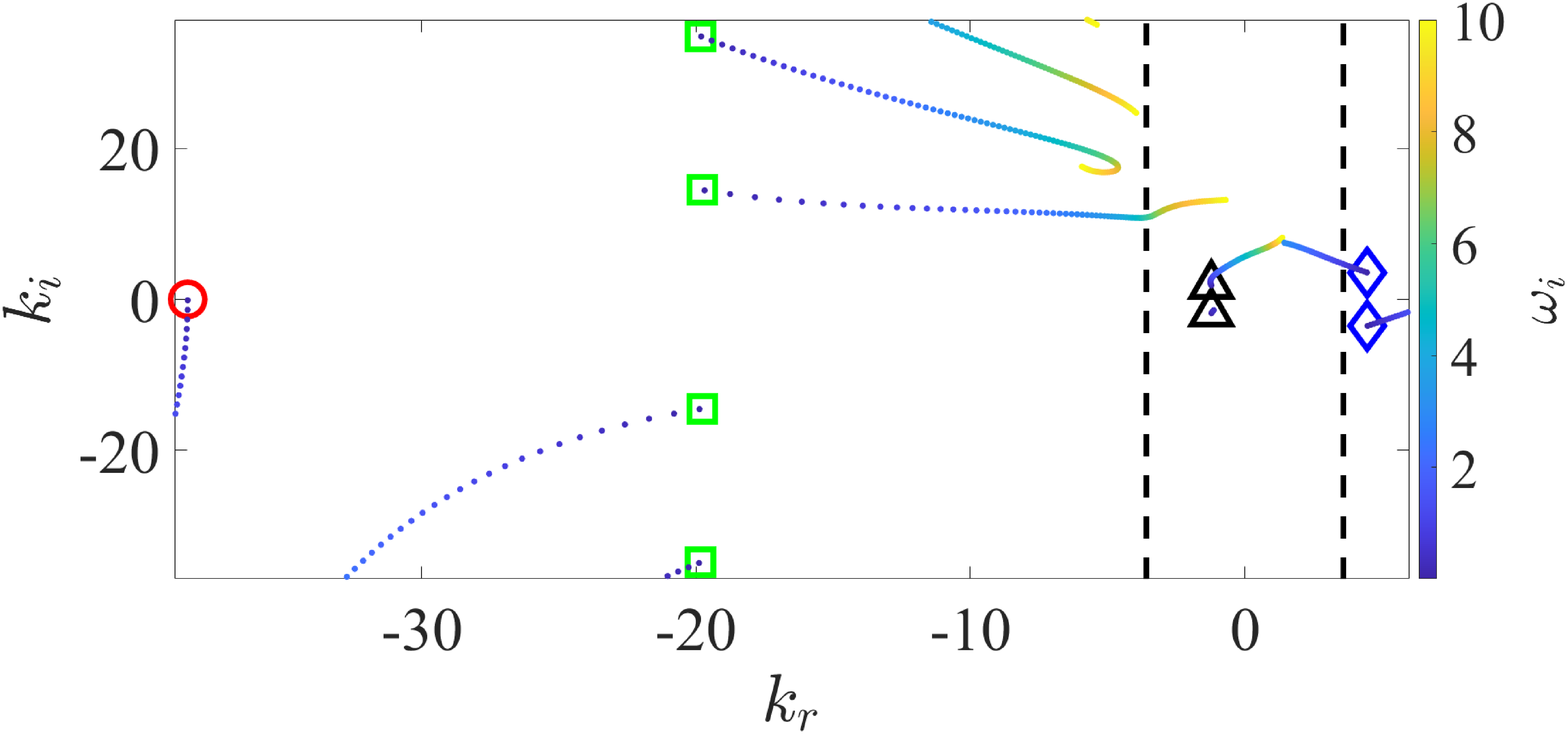}}
\subfigure[]{\includegraphics[scale=0.17]{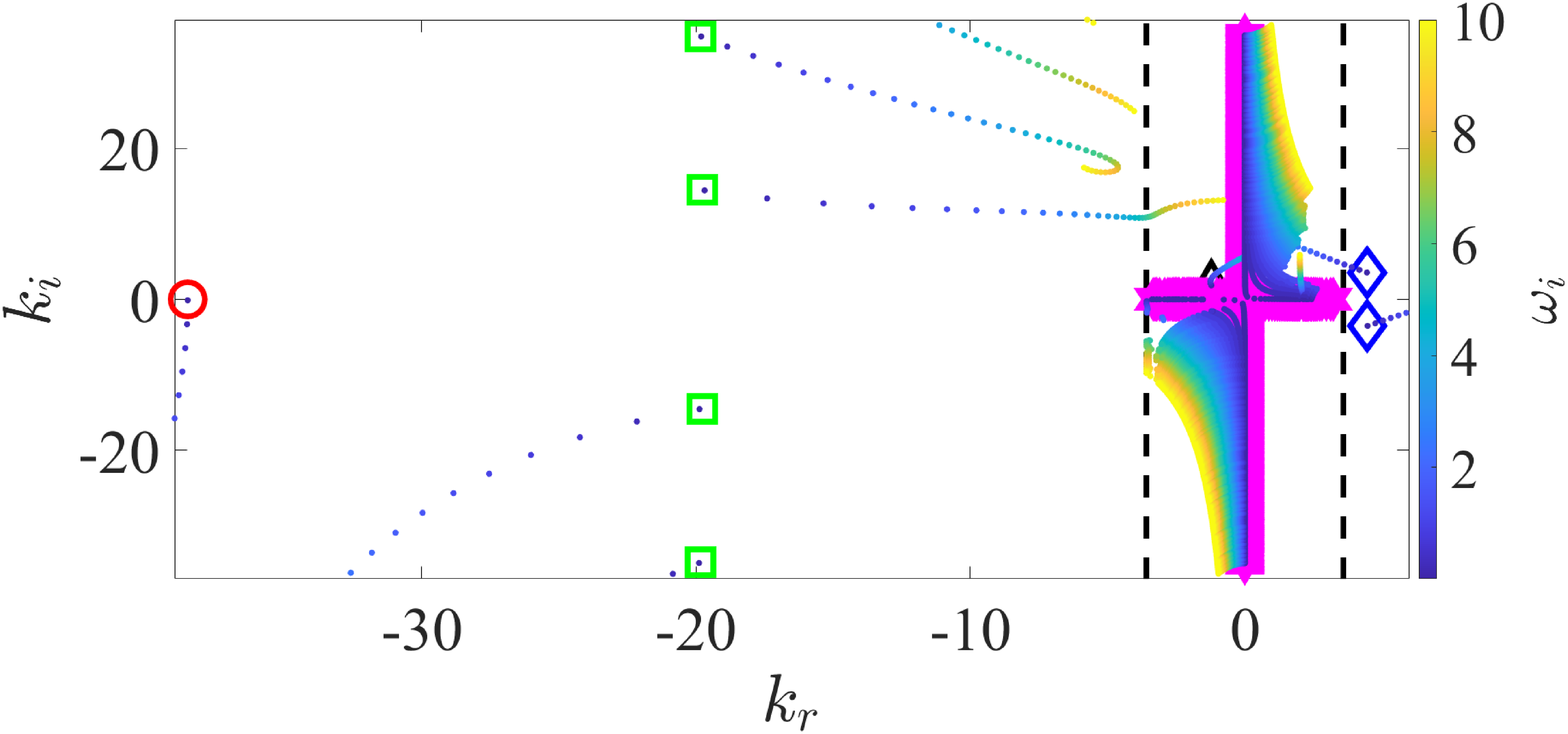}}
\subfigure[]{\includegraphics[scale=0.17]{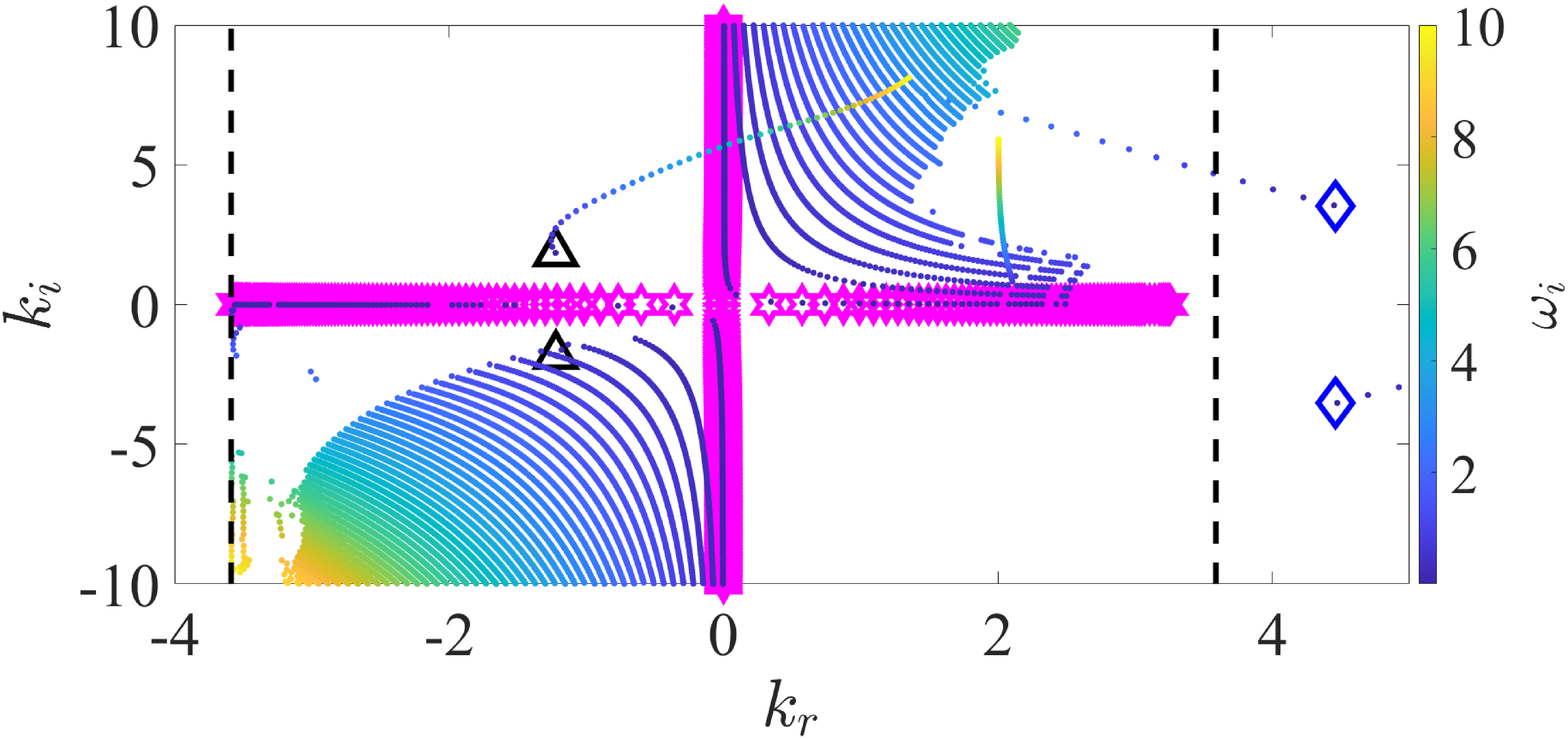}}
\caption{Eigenspectra for $M_j=0.91$, $T\approx 0.85$, $St=0.68$ and $m=0$ obtained from both unconfined and distantly-confined vortex-sheet models. (a) unconfined VS, (b) distantly-confined VS, (c) zoom on the free-stream supersonic region of the distantly-confined V-S spectrum. Markers and colours are the same used in figures \ref{fig:eigenspectrum_standard} and \ref{fig:eigenspectrum_generalised}. Dashed black lines represent the ambient speed of sound $\pm c_\infty$.}
\label{fig:eig_M_effect}
\end{figure}

We here show that the efficiency of the distantly-confined vortex sheet in describing free-stream acoustics along with eigensolutions of the unconfined vortex sheet is independent of the jet-flow condition considered. Figure \ref{fig:eig_M_effect} shows the eigenspectra of both unconfined and distantly-confined vortex-sheet models ($r_{MAX}=100$) for the same frequency considered above, i.e., $St=0.68$, but for subsonic flow conditions, that is, $M_j=0.91$ and $T\approx 0.85$. As outlined above for supersonic flow conditions, the eigenspectrum of the unconfined vortex sheet does not include free-stream acoustic modes, which, on the contrary, do appear in the distantly-confined case. We further note that all the free-jet modes are recovered in the eigenspectrum of the distantly-confined case.

\section{Frequency effect}
\label{app:frequency}
\begin{figure}
\centering
\subfigure[]{\includegraphics[scale=0.18]{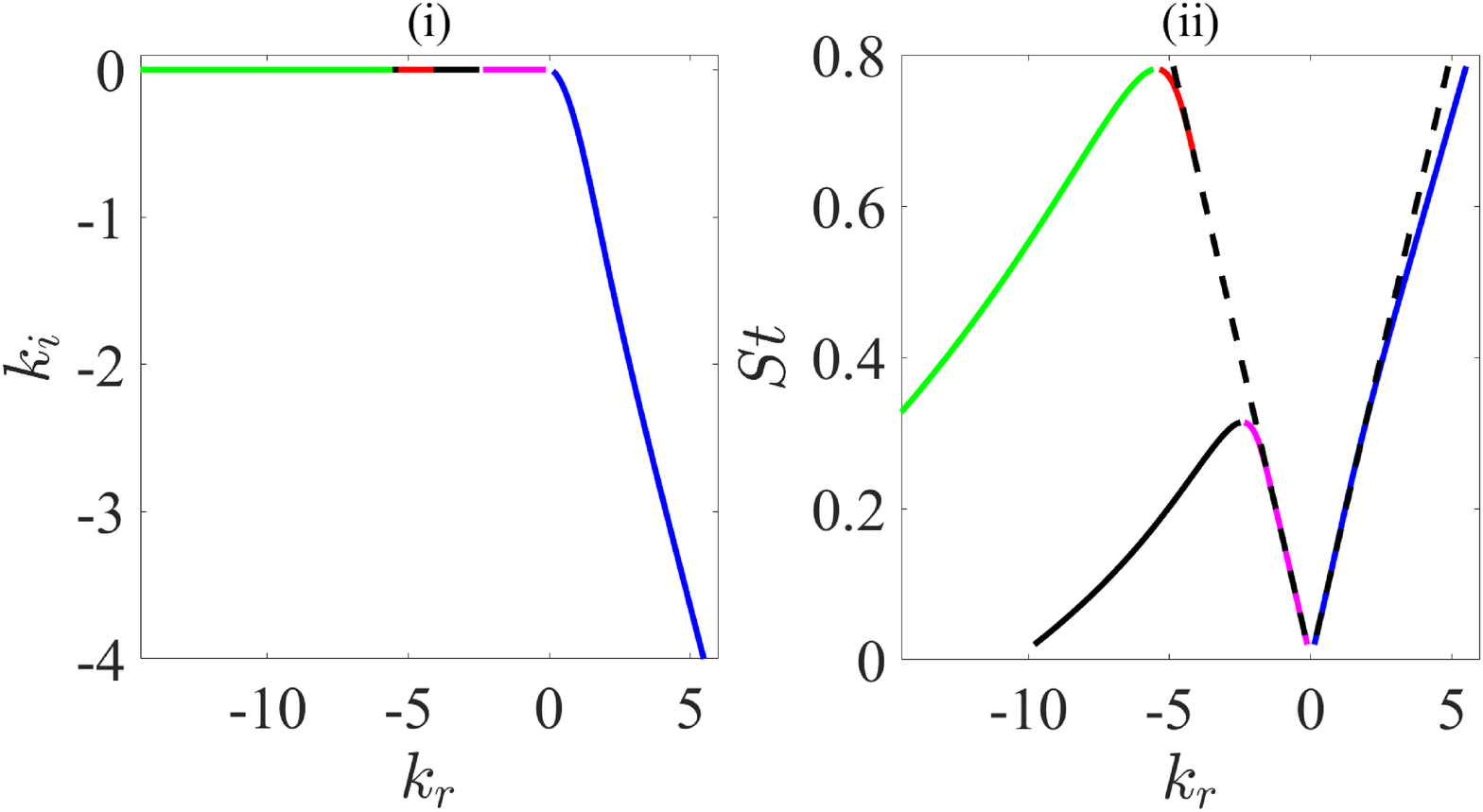}}
\subfigure[]{\includegraphics[scale=0.18]{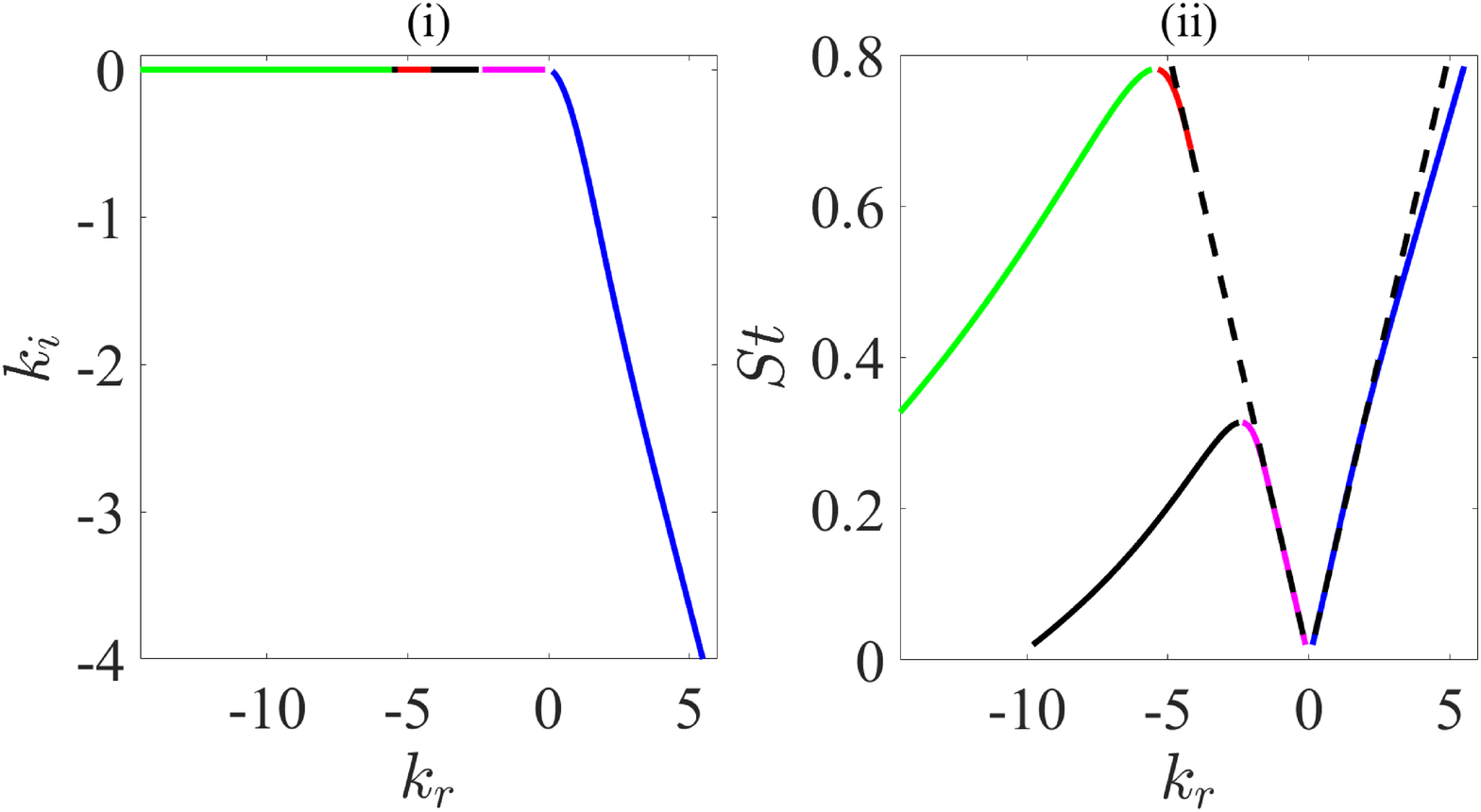}}
\caption{Eigenspectra for $M_j=1.1$, $T\approx 0.81$ and $m=0$ in the obtained from both the unconfined and distantly-confined vortex-sheet models. (a) unconfined VS, (b) distantly-confined VS; (i) $k_r$-$k_i$ plane, (ii) $k_r$-$St$ plane. Colours are the same used in figures \ref{fig:eigenspectrum_standard} and \ref{fig:eigenspectrum_generalised} except for the $k_p^-$ mode with $n=1$ which, for the sake of clarity, is here represented in magenta. Dashed black lines represent the ambient speed of sound $\pm c_\infty$.}
\label{fig:eig_St_effect}
\end{figure}

We here show that the capability of the distantly-confined vortex sheet to describe free unconfined eigensolutions is independent of the frequency considered. For this aim, we report in figure \ref{fig:eig_St_effect} the eigenspectra in the $k_r$-$k_i$ and $k_r$-$St$ planes for the same supersonic flow conditions considered above obtained from both the unconfined and distantly-confined vortex-sheet models. For the sake of brevity, we only show the K-H mode and the downstream- and upstream-travelling guided modes of first and second radial orders and the Strouhal number is varied in the range $\left[0.02,0.8\right]$. No differences between the eigenspectra obtained using the unconfined and the distantly-confined vortex-sheet models can be appreciated in figure \ref{fig:eig_St_effect}.

\bibliography{biblio_rev}

\end{document}